\documentclass[12pt,a4paper]{article}
\usepackage{amsmath}
\usepackage{amsthm}
\usepackage{theorem}
\usepackage{makeidx}
\usepackage{amstext}
\usepackage[cp1251]{inputenc}
\usepackage{epsfig}
\begin{document}
\title{Analytical and numerical solution of coupled KdV-MKdV
system}
\author{A.A. Halim $^{1}$, S.P. Kshevetskii $^{2}$, S.B. Leble$^{3}$ \\
        $^{1,3}$ Technical University of Gdansk, \\
        ul. G. Narutowicza 11/12, 80-952 Gdansk, Poland. \\
        $^{2}$ Kaliningrad state University, Kaliningrad, Russia\\
         $^{3}$leble@mif.pg.gda.pl}
\maketitle
\begin{abstract}
\noindent The matrix 2x2 spectral differential equation of the
second order is considered on x in ($-\infty,+\infty$). We
establish elementary Darboux transformations covariance of the
problem and analyze its combinations.  We select a second
covariant equation to form  Lax pair of a coupled KdV-MKdV system.
The sequence of the elementary Darboux transformations of the
zero-potential seed produce two-parameter solution for the coupled
KdV-MKdV system with reductions. We show effects of parameters on
the resulting solutions (reality, singularity). A numerical method
for general coupled KdV-MKdV system is introduced. The method is
based on a difference scheme for Cauchy problems for arbitrary
number of equations with constants coefficients. We analyze
stability and prove the convergence of the scheme which is also
tested by numerical simulation of the explicit solutions.
\end{abstract}
\section{Introduction}
There are two complementary approaches to integrable systems:
analytical and numerical ones to be developed. Even most profound
analytical IST method cannot give explicit solution of general
Cauchy problem while numeric can, but is rather compulsory in use:
calculations could need powerful computers. May be most
transparent of analytical methods are based on algebraic
structures associated with a problem. To such structures belongs
Darboux transformations covariance of Lax representation of
nonlinear equations that yields a powerful tool for explicit
solutions production. We investigate applications of special kind
of such discrete symmetry - to be called elementary ones
\cite{Leb-Ust}. Its elementarity simply means that a product of
such transformations generate the standard one
\cite{Leb-Ust,Mate}. Here we study combinations of such transforms
that do not coincide with binary ones \cite{Leb-Ust2} and hence
are not so known.

The main ideas of numerical integration of such integrable systems
go up to the famous properties of the equations as the Lax pair
and infinite series of conservation laws existence \cite{Ksh}.
From a point of view of general theory of such systems some hopes
are concerned with a development of the finite-difference or other
approximations of the systems. Namely if one could prove a
convergence and stability theorems for such difference systems
(existence of solutions is implied), a way to existence and
uniqueness of solutions is opened \cite{Lad}.

The coupled KdV-MKdV system arises in many problems of
mathematical physics. Some integrable systems are associated with
a polynomial spectral problem and have Virasoro symmetry algebras
are considered \cite{Wen}. A dispersive system describing a vector
multiplet interacting with the KdV field is a member of a
bi-Hamiltonian integrable hierarchy \cite{Kup}. Recently a
multisymplectic numerical twelve points scheme was produced. This
scheme is equivalent to the multisymplectic Preissmann scheme and
is applied to solitary waves over long time interval \cite{Ping}.
The coupled KdV-MKdV system is also connected to other physical
applications \cite{Leb}.

The general system we consider in this work have the following
form
\begin{equation} \label{KdV-M}
\theta _{t}^{n}+\underset{m,k}{\sum }\left( g_{m,k}^{n,1}\theta
^{m}\theta _{x}^{k}+g_{m,k}^{n,2}(\theta ^{m})^{2}\theta
_{x}^{k}+g_{m,k}^{n,3}\theta _{x}^{m}\theta
_{x}^{k}+g_{m,k}^{n,4}\theta ^{m}\theta
_{xx}^{k}+g_{m,k}^{n,5}\theta ^{m}(\theta ^{k})_{x}^{2}\right)
+d_{n}\theta _{xxx}^{n}=0,
\end{equation}%
where n, m, k =1,2,...N are the dependent variables numbers.
Nonlinear coefficients are $g_{m,k}^{n,l},\,l$ =1,2,..5 and
$d_{n}$ are dispersion coefficients.

In particular, for the system under consideration ($N=3$) the
variables $\theta ^{1},\theta ^{2},\theta ^{3}$ are denoted by f,
u, v to have the integrable system \cite{Leb-Ust}
\begin{gather}\label{gen-sys2}
f_{t}+\frac{1}{2}f_{xxx}+\frac{3}{2}(uf)_{x}-\frac{3}{4}f_{x}f^{2}=0 \notag \\
u_{t}-\frac{1}{4}u_{xxx}-\frac{3}{2}u_{x}u+3vv_{x}+\frac{3}{4}u_{x}f^{2}-%
\frac{3}{2}(f_{x}v)_{x}=0 \notag  \\
v_{t}+\frac{1}{2}v_{xxx}+\frac{3}{2}v_{x}u-\frac{3}{4}(vf^{2})_{x}+\frac{3}{4%
}u_{xx}f+\frac{3}{2}u_{x}f_{x}=0 \label{gen-sys2}
\end{gather}
The Lax pair is given in \cite{Leb-Ust}. The system exhibits two
integrable reductions having explicit solutions, Hirota-Satsuma
\cite{HS, Dodd} and a two components KdV-MKdV system \cite
{Leb-Ust, Leb-Ust2}. Krishnan \cite{Krish} showed that a
generalized KdV-MKdV system have solitary wave solutions and
investigate the effects of increasing the nonlinearity of one
variable on the existence of solitary waves. Some form of KdV-MKdV
system have explicit solutions in terms of Jacobi elliptic
functions \cite{Guha}.

In this work we present explicit solutions for a system of three
equations (\ref{gen-sys2}) that have not been specified in
\cite{Leb-Ust,Leb-Ust2}. We study this two-parameter explicit
solutions and show effects of choosing these parameters on the
solutions. We demonstrate the use of two arbitrary elementary DTs
\cite{Leb-Ust} and its special choice that holds a hereditary of
the reduction to built explicit solutions to the KdV-MKdV system
(\ref{gen-sys2}).

Also we modify a numerical method \cite{Ksh,Halim} for solution of
system (\ref{KdV-M}). It is a difference scheme for Cauchy
problems for arbitrary number of equations with constants
coefficients. The scheme preserves two conservation laws for the
KdV type equations and the order of error of the difference
formulas is improved \cite{Ksh,Zhu}. The convergence is proved and
stability is analyzed giving the conditions taken in account in
choosing time and space step sizes \cite{And,Lan} .

The present work is organized as follows. Section 2 introduces the
matrix spectral equation of the second order with 2 x 2 matrix
coefficients and two elementary DTs. We select the second equation
of the Lax pair and derive the compatibility conditions. The
product of these two transformations yields the standard DT
\cite{Leb-Ust,Mate}. Section 3 illustrates how the, first,
elementary DT is used to produce solution to the KdV equation as
well as the general evolution equation generated by the
compatibility conditions of the Lax pair. Explicit solutions are
introduced for the case of zero initial potentials of the matrix
problem. In section 4 we consider a reduction constraints on the
potential of the matrix spectral equation. This reduction gives an
automorphism that relates two pairs of solution of the spectral
equation for two spectral parameters. We use this results in the
compound elementary DTs to produce an explicit solution to a
coupled KdV-MKdV system that results from the compatibility
conditions of Lax pair under this reduction. The effects of these
parameters on the solution (reality, singularity) is analyzed.
Section 5 introduces a numerical method for solving coupled
KdV-MKdV system (\ref{KdV-M}). We produce a difference scheme for
a Cauchy problem with initial condition rapidly decreasing at both
infinities. The main steps of the scheme convergence and stability
analysis is shown while the details are explained in appendix A
and B. The scheme is tested by applying it to integrable coupled
KdV-MKdV system and the numerical results are compared with
explicit formulas obtained in section 4.
\section{Lax pair spectral equations and the elementary DTs }
Consider a matrix spectral equation of the second order with
spectral parameter $\lambda$ and $2\times2$ matrix coefficients.
\begin{equation}  \label{LP 1}
\Psi_{xx}+F \Psi_{x}+U \Psi=\lambda \sigma_{3} \Psi
\end{equation}
where the vector $\Psi= \left(\Psi_{1}, \Psi_{2}\right)^{T}$and
the matrix
potentials are $U=\{u_{ij}\},F=\{f_{ij}, f_{ii}=0\}, i=1,2$ while $%
\sigma_{3}=diag(1, -1)$ is the Pauli matrix.\newline

For equation (\ref{LP 1}) we perform two elementary Darboux transforms \cite%
{Leb-Ust}. The first one is
\begin{gather}\label{1DTA}
\overset{\sim }{\Psi }_{1}=\Psi _{1x}+\epsilon _{11}\Psi
_{1}+\epsilon_{12}\Psi _{2}\,,\,\ \ \ \epsilon _{11}=-\left( \varphi _{1x}+\frac{1}{2}%
f_{12}\varphi _{2}\right) /\varphi _{1}\,,\,\ \epsilon _{12}=f_{12}/2, \notag \\
  \overset{\sim }{\Psi }_{2}=\Psi _{2}+\ \epsilon _{21}\,\Psi_{1}\,,\ \ \ \ \
\ \ \ \ \ \ \ \ \ \ \ \ \ \epsilon _{21}=-\ \varphi _{2}/\varphi _{1} , \notag \\
  \overset{\sim }{\varphi }_{1}=\left( \partial _{x}+\epsilon_{11}\right)
\varphi _{3}+\epsilon _{12}\,\varphi _{4}\,\,,\,\ \overset{\sim }{\varphi }%
_{2}=\varphi _{4}+\epsilon _{21}\,\varphi _{3}. \label{1DTA}
\end{gather}
where $\left( \varphi _{1},\,\varphi _{2}\right) ^{T}\,\ $ and
$\left(\varphi _{3},\,\varphi _{4}\right) ^{T}\,\ $ are two solutions of (\ref{LP 1}%
) corresponding to different spectral parameters.

Substituting the above expressions for $\overset{\sim }{\Psi }_{1},\,\overset%
{\sim }{\Psi }_{2}\,$into (\ref{LP 1}) and collecting the coefficients of $%
\Psi _{1}$,\thinspace $\Psi _{2}\,\ $and their derivatives we
obtain the expressions for new potentials as
\begin{gather}\label{1DTB}
\overset{\sim }{f}_{12}=u_{12}+f_{12}\,\,\epsilon _{11}, \notag\\
\overset{\sim }{f}%
_{21}=-2\,\epsilon _{21}\smallskip , \notag \\
\overset{\sim }{u}_{11}=u_{11}-2\epsilon _{11x}-\overset{\sim }{f}%
_{12}\,\epsilon _{21}-f_{21}\,\epsilon _{12}\smallskip, \smallskip \notag \\
\overset{\sim }{u}_{12}=u_{12x}-\epsilon _{12xx}+\epsilon
_{11}u_{12}-\epsilon _{12}\left( \overset{\sim
}{u}_{11}+u_{22}\right)
\smallskip, \notag\\
\overset{\sim }{u}_{21}=f_{21}-2\epsilon _{21x}-\overset{\sim }{f}%
_{21}\epsilon _{11}\smallskip,  \notag\\
\overset{\sim }{u}_{22}=u_{22}-\epsilon _{21}u_{12}-\overset{\sim }{u}%
_{21}\epsilon _{12}-\overset{\sim }{f}_{21}\epsilon _{12x}.
\label{1DTB}
\end{gather}
The second elementary DT is performed after the first one and can
be obtained by reversing the indices 1$\rightarrow 2\,$and
2$\rightarrow 1\,\ $to get, for example\smallskip , the following
potentials
\begin{gather}\label{2DT}
\overset{\approx }{f}_{21}\,=\overset{\sim }{u}_{21}+\overset{\sim }{f}_{21}%
\overset{\sim }{\epsilon }_{22},{\hspace{2 mm}} \overset{\sim }{\epsilon }%
_{22}=-\left( \overset{\sim }{\varphi }_{2x}+\frac{1}{2}\overset{\sim }{f}%
_{21}\overset{\sim }{\varphi }_{1}\right) \,/\overset{\sim }{\varphi }_{2} , %
 \notag \\
\overset{\approx }{u}_{22}=\overset{\sim }{u}_{22}-2\overset{\sim
}{\epsilon
}_{22x}-\overset{\approx }{f}_{21}\overset{\sim }{\epsilon }_{12}-\overset{%
\sim }{f}_{12}\overset{\sim }{\epsilon
}_{21},\,\smallskip{\hspace{3 mm}}
\overset{\sim }{\epsilon }_{12}=-\ \overset{\sim }{\varphi }_{1}/\overset{%
\sim }{\varphi }_{2},{\hspace{3 mm}} \overset{\sim }{\epsilon }_{21}=\overset%
{\sim }{f}_{21}/\,2 .\ \label{2DT}
\end{gather}
The spectral equation (\ref{LP 1}) is considered as the first
equation of the Lax pair, take the second as
\begin{equation}  \label{LP 2}
\Psi_{t}=\Psi_{xxx}+B \Psi_{x}+C \Psi ,
\end{equation}
where $B=\frac{3}{2} diagU+\frac{3}{2} F_{x}+\frac{3}{4}
F^{2}$\newline
and $C=\frac{3}{2} U_{x}-\frac{3}{4}diagU_{x}-\frac{3}{4}%
(f_{12}u_{21}+f_{21}u_{12})I+ \frac{3}{8}(f_{12,x}f_{21}-f_{12}f_{21,x})%
\sigma_{3}+ \frac{3}{4}(u_{11}-u_{22})\sigma_{3}F.$\\
\,\,\\
Equation (\ref{LP 2}) is also covariant under transformations
(\ref{1DTA}), (\ref{1DTB}). The compatibility conditions have the
following form
\begin{gather} \label{compat}
F_{t}-F_{3x}+B_{2x}-3U_{2x}+2C_{x}+FB_{x}- \sigma_{3}B
\sigma_{3}F_{x}
+UB \notag  \\
- \sigma_{3}B \sigma_{3}U+FC
- \sigma_{3}C \sigma_{3}F=0, \notag  \\
U_{t}-U_{3x}+C_{2x}+UC- \sigma_{3}C \sigma_{3}U+FC_{x} -
\sigma_{3} B \sigma_{3}U_{x}=0 .
\end{gather}
and the transformations (\ref{1DTA}), (\ref{1DTB}) determine a
discrete symmetry of (\ref{compat})
\section{Solution of two coupled KdV-MKdV equations and KdV equation via the
first elementary \thinspace DT}

For a spectral parameter $\lambda $ and a seed potential F, U we
obtain the solutions $\varphi_{1}, \varphi_{2}$ to the pair
(\ref{LP 1}), (\ref{LP 2}). Then performing the first elementary
DT to obtain
the new potentials $ \overset{\sim}{F}, \overset{\sim}{U}$ which are solutions to the system (%
\ref{compat}). For the case of zero seed potential the solutions, $%
\varphi _{1}$ and  $\varphi _{2}$ of the system $  (\ref{LP 1}),
(\ref{LP 2})$ have the form
\begin{gather} \label{fi}
\varphi _{1}=c_{1}e^{a x + a ^{3} t}+c_{2}e^{-(a x + a ^{3} t)}, \notag\\
\varphi _{2}=d_{1}e^{i a x + ( i a) ^{3} t }+d_{2}e^{-(i a x + (i
a) ^{3} t )} . \label{fi}
\end{gather}%
where $c_{1},\,c_{2},\,d_{1},\,\,d_{2}\,\ $are arbitrary
constants, $a= \sqrt{\lambda} $ and i is the imaginary unit.
System (\ref{compat}) reduced (for the only nonzero elements ) to
the following
\begin{gather}\label{gen-sys}
f_{21t}+\frac{1}{2} f_{21xxx}+\frac{3}{4} f_{21}u_{11x} =-\frac{3}{2}%
u_{11}u_{21},  \notag \\
u _{11t}-\frac{1}{4}u_{11xxx}-\frac{3}{2}u_{11}u_{11x}=0, \notag   \\
u _{21t}+\frac{1}{2}u_{21xxx}+\frac{3}{4}%
u_{21}u_{11x}+\frac{3}{2}u_{21x}u_{11} =\frac{3}{4}\
u_{11}f_{21xx} +\frac{3}{4}u_{11}^{2}f_{21}. \label{gen-sys}
\end{gather}
where $f_{12}=0, u_{12}=0, u_{22}=0$ and tildes are omitted for
simplicity. This system with explicit solution obtained from
(\ref{1DTA}), (\ref{1DTB}) as
\begin{gather} \label{sol-gen-sys}
{f}_{21}=\left( 2e^{\left( 1-i\right) a\left( a^{2}t+x\right)
}\left( d_{2}e^{2ia^{3}t}+d_{1}e^{2iax}\right) \right) /\left(
c_{2}+c_{1}e^{2a\left( a^{2}t+x\right) }\right),  \notag  \\
{u}_{11}=\left( 8a^{2}c_{1}c_{2}e^{2a\left( a^{2}t+x\right)
}\right) /\left(
c_{2}+c_{1}e^{2a\left( a^{2}t+x\right) }\right)^{2},  \notag  \\
{u}_{21}=\left( -2iae^{\left( 1-i\right) a\left( a^{2}t+x\right)
}\left( d_{2}e^{2ia^{3}t}-d_{1}e^{2iax}\right) \right) /\left(
c_{2}+c_{1}e^{2a\left( a^{2}t+x\right) }\right).
\label{sol-gen-sys}
\end{gather}
where $c_{1},\,c_{2},\,d_{1},\,\,d_{2}\,\ $are arbitrary
constants and $a= \sqrt{\lambda}$.\newline

The second equation in (\ref{gen-sys}) is the KdV equation while
the remaining are a two components coupled KdV-MKdV system that
was solved by elementary DT.

\section{Solution of three coupled KdV-MKdV equations via the compound
elementary DTs}

Existence of different kinds of automorphism causes special
constraints \cite{Leb-Ust}. Multiplying (\ref{LP 1}) by $\sigma
_{1}=
\begin{pmatrix}
0 & 1 \\
1 & 0%
\end{pmatrix}%
$ to have
\begin{equation}  \label{Auto1}
\sigma _{1}\Psi_{xx}+\sigma _{1} F \Psi _{x}+\sigma _{1} U
\Psi=\lambda \sigma _{1}\sigma _{3}\Psi
\end{equation}
but $\sigma _{1}\sigma _{3}=-\sigma _{3}\sigma _{1} $ and consider
the conditions $\sigma_{1} F= F \sigma_{1}$ and $\sigma_{1} U= U
\sigma_{1}$ that means
\begin{equation}  \label{constr}
f_{12}=f_{21}=f,\hspace*{5 mm} u_{11}=u_{22}=u, \hspace*{5
mm}u_{12}=u_{21}=v.
\end{equation}
So (\ref{Auto1}) becomes
\begin{gather*}
\left( \sigma _{1}\Psi \right) _{xx} + F \left(\sigma
_{1}\Psi\right) _{x}+U \left(\sigma _{1} \Psi\right)=-\lambda
\sigma _{3}\left( \sigma _{1}\Psi \right)
\end{gather*}

The above automorphism $\Psi \left( \lambda \right) \leftarrow
\sigma
_{1}\Psi \left( -\lambda \right) $ relates two pair of solutions $%
(\varphi_{1}, \varphi_{2})$ and $(\varphi_{3}, \varphi_{4})$ of
(\ref{LP 1}) corresponding to different values of spectral
parameter $\lambda , -\lambda $ as\newline
\begin{gather*}
\begin{pmatrix}
\varphi_{3}(-\lambda) \\
\varphi_{4}(-\lambda)%
\end{pmatrix}%
=\sigma_{1}%
\begin{pmatrix}
\varphi_{1}(\lambda) \\
\varphi_{2}(\lambda)%
\end{pmatrix}%
=%
\begin{pmatrix}
\varphi_{2}(\lambda) \\
\varphi_{1}(\lambda)%
\end{pmatrix}%
\end{gather*}

Using this result in the elementary DTs (\ref{1DTA}), (\ref{1DTB}) and (\ref%
{2DT}) to obtain the expressions for the new potentials f, u, v.
In the case of zero initial potentials these new potentials have
the following forms
\begin{gather}\label{fuv1}
f= 2 \frac{\varphi _{1} (\varphi _{2})_{x}-\varphi _{2} (\varphi
_{1})_{x}}{(\varphi _{1})^{2}-(\varphi _{2})^{2}},\notag
\\\newline
u= \left(\frac{(\varphi _{1}^{2})_{x}-(\varphi _{2}^{2})_{x}}
{(\varphi _{1})^{2}-(\varphi
_{2})^{2}}\right)_{x}+2\left(\frac{\varphi _{1} (\varphi
_{2})_{x}-\varphi _{2}
(\varphi _{1})_{x}}{(\varphi _{1})^{2}-(\varphi _{2})^{2}}\right)^{2}, \notag \\
v= 2 \left(\frac{\varphi _{1} (\varphi _{2})_{x}-\varphi _{2}
(\varphi _{1})_{x}}{(\varphi _{1})^{2}-(\varphi
_{2})^{2}}\right)_{x}+\frac{\left(\varphi _{1}(\varphi
_{2})_{x}-\varphi _{2}(\varphi _{1})_{x}\right)\left((\varphi
_{1}^{2})_{x}-(\varphi _{2}^{2})_{x}\right)}{ \left( (\varphi
_{1})^{2}-(\varphi _{2})^{2} \right)^{2}}.
\end{gather}
where $\varphi_{1}, \varphi_{2}$ are as in (\ref{fi}) with $%
c_{1},\,c_{2},\,d_{1},\,\,d_{2}\,\ $are arbitrary constants and $a= \sqrt{%
\lambda}$.\newline The above expressions are solutions of system
(\ref{compat}) that reduced under the reduction conditions
(\ref{constr}) to system (\ref{gen-sys2}).

The choice of the arbitrary constants
($c_{1},\,c_{2},\,d_{1},\,\,d_{2}$) affects on the behavior of
the solution in formula (\ref{fuv1}). For example choosing equal
constants $c_{1}=c_{2}=d_{1}=d_{2}=0.5$ (we choose the value to
be $0.5$ to simplify the resulting formula but the idea valid for
any value) the solutions have the form
\begin{gather}\label{fuv2}
f=2 a (sin \eta_{1} cosh\eta_{2}-cos\eta_{1}
sinh\eta_{2})/(cosh^{2}
\eta_{2}-cos^{2} \eta_{1}),  \notag  \\
u=2 a^{2}(sin \eta_{1}
cosh\eta_{2}+cos\eta_{1}sinh\eta_{2})^{2}/(cosh^{2}
\eta_{2}-cos^{2} \eta_{1})^{2} , \notag \\
v=2a^{2}(cos3\eta_{1}cosh\eta_{2}-2sin \eta_{1}sinh\eta_{2}(cos2
\eta_{1}+cosh2 \eta_{2}+2)-cos\eta_{1}cosh3 \eta_{2})   \notag \\
\hspace*{5 mm}/(cosh^{2}\eta_{2}-cos^{2} \eta_{1})^{2}.
\label{fuv2}
\end{gather}
where $\eta_{1}=a^{3}t-ax, \hspace*{3 mm} \eta_{2}=a^{3}t+ax, a=\sqrt{\lambda%
}$ is real.
\,\\

We see that the above expression (\ref{fuv2}) is singular at
$\eta_{2}=0,
\eta_{1}=n\pi, n=0,1,2,...$. Hence we have singularity at $(x=\frac{n\pi}{2a}%
, t=\frac{n\pi}{2a^{3}})$.\newline

To obtain continuous solutions we can choose $c_{1}= c_{2},
d_{1}= d_{2}= r . c_{1}$, r is real constant. We again choose
$c_{1}=0.5 $ following the previous concept. So (\ref{fuv1}) have
the form
\begin{gather} \label{sol-gen-sys2}
f= 2ar\left( \cosh \eta _{2}\sin \eta _{1}-\cos \eta _{1}\sinh
\eta _{2}\right) /\left( \cosh ^{2}\eta _{2}-r^{2}\cos ^{2}\eta
_{1}\right) , \notag  \\
u= a^{2}\left( 1-r^{4}-r^{4}\cos 2\eta _{1}+\cosh 2 \eta
_{2}+r^{2}\sin 2\eta _{1}\sinh 2\eta _{2}\right)
 /\left( \cosh^{2}\eta _{2}-r^{2}\cos ^{2} \eta
_{1}\right) ^{2} , \notag  \\
v= 2a^{2}r(((-7+6r^{2}+2r^{2}\cos 2 \eta
_{1})\cos \eta _{1} \cosh \eta _{2}-\cos \eta _{1} \cosh 3 \eta _{2}) \notag \\
-2(1+r^{2}+r^{2}\cos 2 \eta _{1}+\cosh 2 \eta _{2})  \sin \eta
_{1}\sinh \eta _{2})) / (-1+r^{2}+r^{2}\cos 2 \eta _{1}-\cosh 2
\eta _{2}) ^{2}
\end{gather}
where $a, \eta_{1}, \eta_{2}$ as in (\ref{fuv2})
\,\,\\\

Choosing this parameter ($r$) to be $r$ $<1$ gives real
nonsingular solutions. The above formula (\ref{sol-gen-sys2}) is
built from elliptic and periodic functions so it does not preserve
its symmetry but its localized as shown in figures (1.a,b) below.
\begin{center}
\epsfig{file=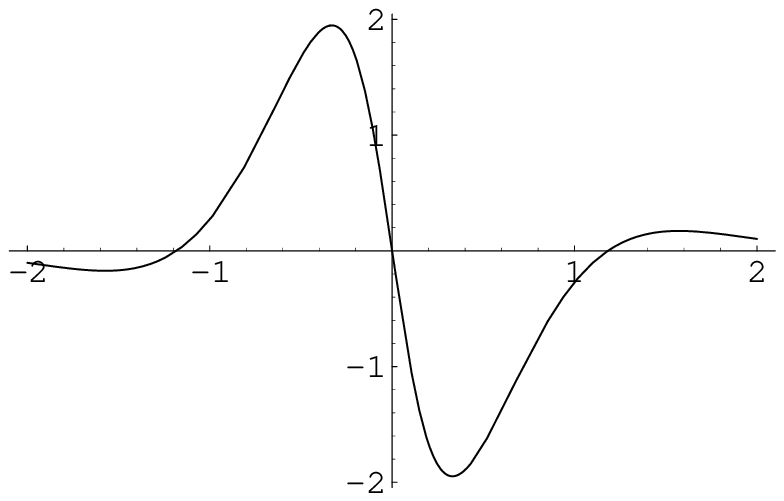, height=2.5 cm, width=4.5 cm ,clip= ,
angle=0} \epsfig{file=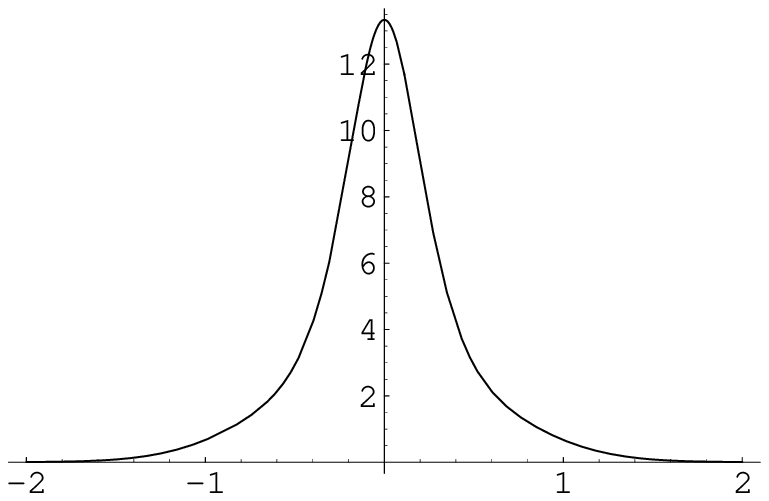,height=2.5 cm, width=4.5cm
,clip=,angle=0} \epsfig{file=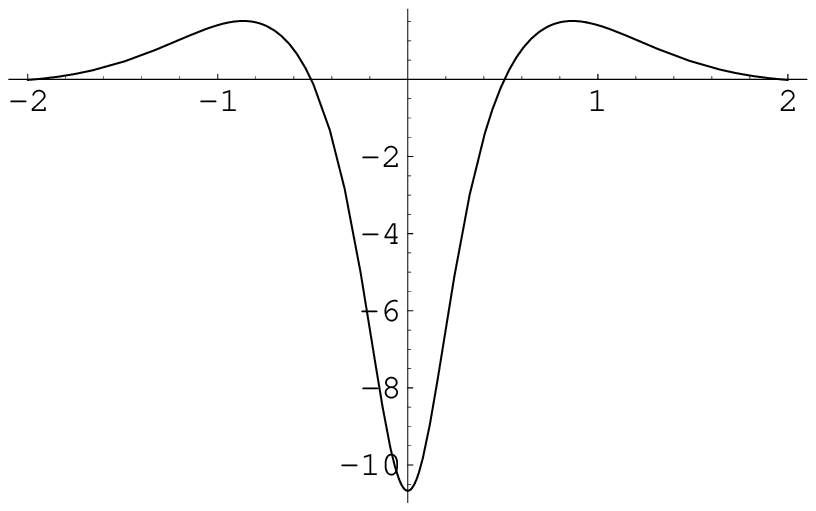,height=2.5cm, width=4.5cm
,clip=,angle=0}\\
\ Fig.(1.a) Non-singular solutions, f, u and v (r=0.5) , a=2,
t=0.\\
\epsfig{file=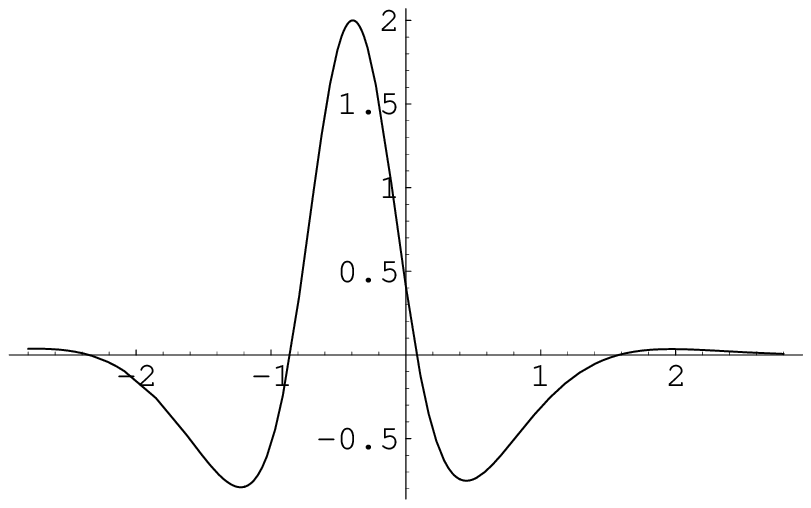, height=2.5 cm, width=4.5 cm ,clip= ,
angle=0} \epsfig{file=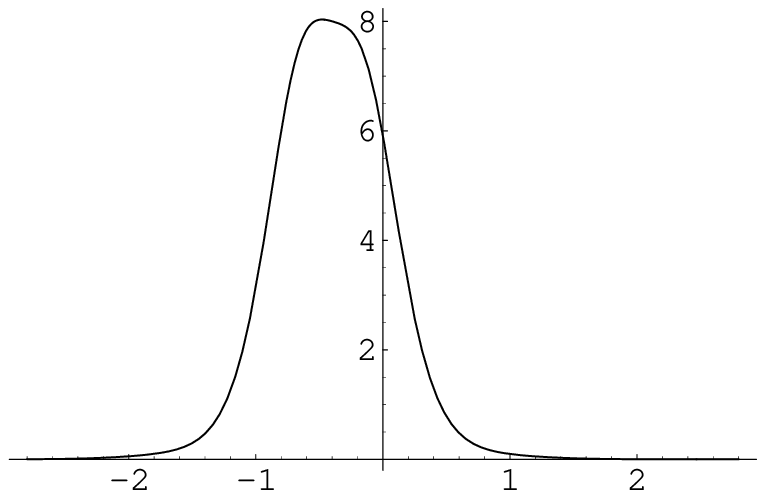,height=2.5 cm, width=4.5cm
,clip=,angle=0} \epsfig{file=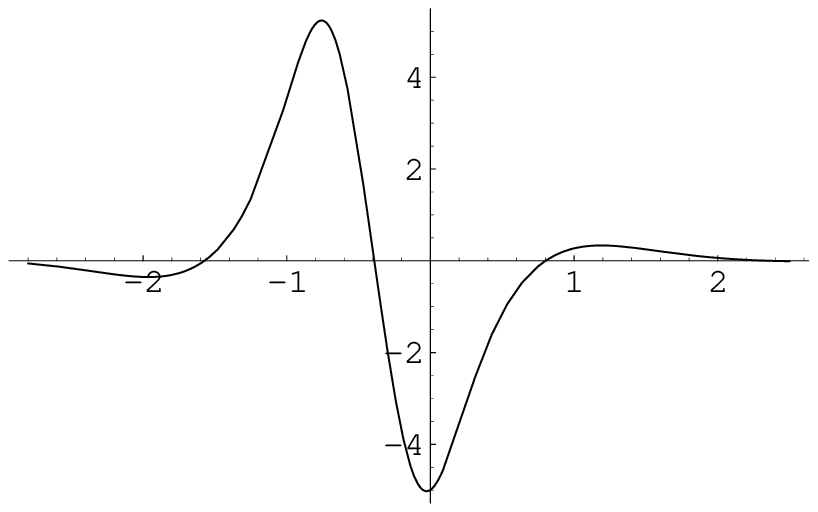,height=2.5cm,
width=4.5cm
,clip=,angle=0}\\[0pt]
\ Fig.(1.b) Propagation of solutions, f, u and v (r=0.5) , a=2,
t=1.\\
Fig.(1) The solutions in (\ref{sol-gen-sys2}) does not preserve
its symmetry but its localized.
\end{center}

 Choosing the parameter r to be $r$ $>1$ in formula (\ref{sol-gen-sys2}) gives singular solutions as shown figure
(2) below.
\begin{center}
\epsfig{file=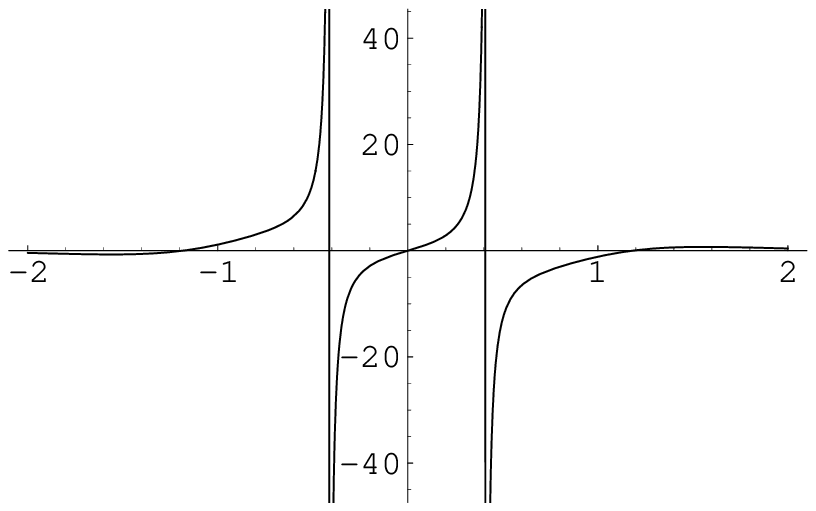, height=2.5 cm, width=4.5 cm ,clip= ,
angle=0} \epsfig{file=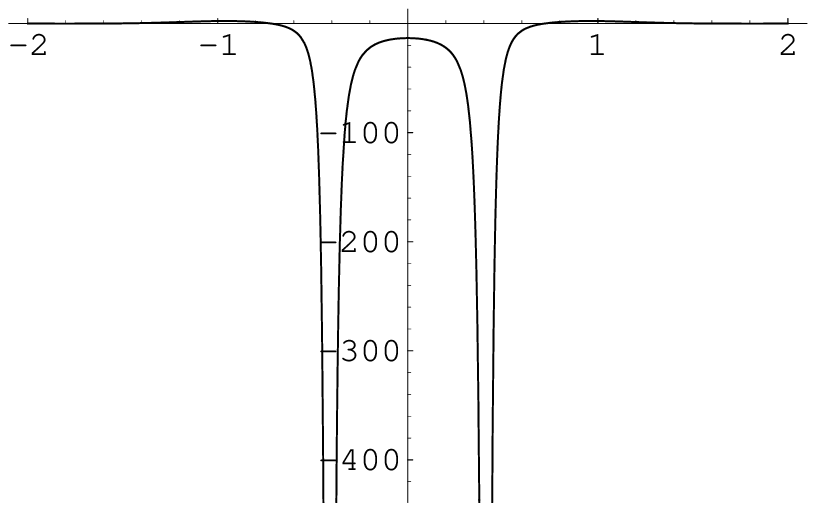,height=2.5 cm, width=4.5cm
,clip=,angle=0} \epsfig{file=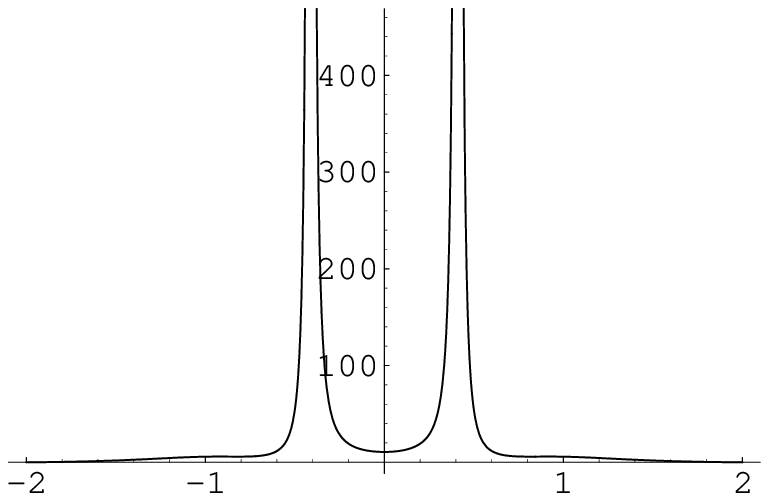,height=2.5 cm,
width=4.5cm
,clip=,angle=0}\\[0pt]
\ Fig.(2) Singular solutions, f, u and v (r=2), a=2, t=0.\\
\end{center}
Moreover the choice of these arbitrary constants ($c_{1},\,c_{2},\,d_{1},\,%
\,d_{2}\,\ $) as well as the spectral parameter $\lambda$ affects
the reality of the resulting solution. As example for
$\lambda=-2im^{2}$, m is real and choosing
$c_{1}=c_{2}=d_{1}=d_{2}=0.5$, we get real solution
\begin{equation*}
f = m (cos2 \zeta_{1} sinh\zeta_{2}-sinh\zeta_{2}-sin \zeta_{1}
cosh2\zeta_{2} +sin\zeta_{1})/(0.25 cosh2\zeta_{2}-0.25)(1-cos2
\zeta_{1})
\end{equation*}
where $\zeta_{1}=2mx+4m^{3}t,\hspace*{2 mm}
\zeta_{2}=2mx-4m^{3}t$, while choosing $c_{1}=c_{2}=1,
d_{1}=d_{2}=2$ give the following complex solution
\begin{gather*}
f=m*(-8(-5sinh\zeta_{2}cos2\zeta_{1}+5sinh\zeta_{2}
+5sin\zeta_{1}cosh2\zeta_{2}-5sin\zeta_{1})-8i(6sinh\zeta_{2}\\
+3sin2\zeta_{1}cosh\zeta_{2}
+3cos\zeta_{1}sinh2\zeta_{2}+6sin\zeta_{1})) /
(17cosh2\zeta_{2}+10+36cos\zeta_{1}cosh\zeta_{2}\\
-8cos2\zeta_{1}cosh2\zeta_{2}+17cos2\zeta_{1} )
\end{gather*}
\section{The numerical method}
\subsection{The difference scheme}

For the coupled KdV-MKdV system (\ref{KdV-M}) we introduce a
numerical (finite difference) method of solution \cite{Ksh, Zhu}.
This scheme is valid for arbitrary number of equations with
constants coefficients and of the form
\begin{gather}\label{schem}
\frac{\theta _{i}^{n,j+1}-\theta _{i}^{n,j}}{\tau
}+\underset{m,k}{\sum } ( g_{m,k}^{n,1}\theta
_{i}^{m,j}\frac{\theta _{i+1}^{k,j}-\theta _{i-1}^{k,j}}{
2h}+g_{m,k}^{n,2}(\theta _{i}^{m,j})^{2}\frac{\theta
_{i+1}^{k,j}-\theta _{i-1}^{k,j}}{2h} \notag \\
+g_{m,k}^{n,3}\frac{\theta _{i+1}^{m,j}-\theta _{i-1}^{m,j}
}{2h}\frac{\theta _{i+1}^{k,j}-\theta _{i-1}^{k,j}}{2h}
+g_{m,k}^{n,4}\theta _{i}^{m,j}\frac{\theta _{i+1}^{k,j}-2\theta
_{i}^{k,j}+\theta _{i-1}^{k,j}}{2h} \notag
\\
+g_{m,k}^{n,5}\theta _{i}^{m,j}\theta _{i}^{k,j}\frac{\theta
_{i+1}^{k,j}-\theta _{i-1}^{k,j}}{2h} )
 +d_{n}\frac{ \theta
_{i+2}^{n,j}-2\theta _{i+1}^{n,j}+2\theta _{i-1}^{n,j}-\theta
_{i-2}^{n,j}}{2h^{3}}=0 \label{schem}
\end{gather}
where i and j are the discrete space and time respectively. The
time step is denoted by $\tau \,$while h denotes spatial step.

\subsection{Stability analysis of the scheme}

We prove stability with respect to small perturbations of initial
conditions \cite{And, Lan}. It is the boundness of the discrete
solution with respect to small perturbation of the initial data.
We give here the main steps while the details are presented in
Appendix A. We can write

\begin{gather*}
d\theta _{i}^{n,j+1}=T_{i,r}^{n,j+1}\,d\theta
_{r}^{n,j}=T_{i,r}^{n,j+1}\,T_{i,r}^{n,j}\,d\theta _{r}^{n,j-1}=\underset{r}{%
\Pi }\,\left( T_{i,r}^{n}\right) ^{r}\,d\theta _{r}^{n,o}
\end{gather*}

where $d\theta _{i}^{n,j+1}$ is perturbations of the discrete solution, $%
d\theta _{r}^{n,o}$ small perturbation of the initial data and $%
T_{i,r}^{n,j+1}$ is a differentiable operator. Stability required
the boundedness of \thinspace $\underset{r}{\Pi }\,\left(
T_{i,r}^{n}\right) ^{r}\,\,$i.e\thinspace\ $\left\| T^{r}\right\|
\,$is bounded.

We found that
\begin{gather} \label{a}
\left\| T^{j+1}\right\| ^{2}\leq e^{a(\tau,h) \tau }, \notag  \\
a(\tau,h) = 2 \underset{l,n,m,k}{\max }\left| g_{m,k}^{n,l}\right| \,\ \underset{%
i,m,k}{\max }\left( \left| \theta _{x,i}^{m,j}\right| \,\left|
\theta _{x,i}^{k,j}\right| \right) +\tau [
\underset{l,n,m,k}{\max }\left| g_{m,k}^{n,l}\right| \,\
\underset{i,m,k}{\max }\left( \left| \theta _{x,i}^{m,j}\right|
\,\left| \theta _{x,i}^{k,j}\right| \right) \notag \\
+\frac{1}{h} \,\underset{l,n,m,k}{\max }\left|
g_{m,k}^{n,l}\right| \,\ \underset{i,m,k}{ \max }\left( \left|
\theta _{i}^{m,j}\right| \,\left| \theta _{i}^{k,j}\right|
\right) +\frac{3\,}{h^{3}}\underset{n}{\max }\left| d_{n}\right|
]^{2} \label{a}
\end{gather}%

The scheme is stable if $a(\tau,h)\leq const$. We have here a
conditional stability. That is we require that $\tau \rightarrow
0\,\ $ more faster than $h\rightarrow 0\,$. Namely we need

\begin{gather*}
\tau \leq \,\left( \text{constant} \right) \,. \,h^{6}
\end{gather*}

\subsection{Convergence proof for the scheme}

We prove that the solution of (\ref{schem}) converges to the solution of (%
\ref{KdV-M}) if the exact solution is continuously differentiable one \cite%
{And, Lan}. We introduce here the main points for the scheme
convergence and give the details in Appendix B.

$\theta _{i}^{j}\,$\ is the difference solution of (\ref{schem}), u$%
_{i}^{j}\,$\ is the exact solution. Hence the error\ v$_{i}^{j}$
is given by $\ \ v_{i}^{j}=\theta _{i}^{j}-u_{i}^{j}\,\ $.
Introducing L$_{2}\ $norm defined by $\left\| V^{j}\right\|
=\left( \underset{i}{\sum }\underset{n}{\sum }\left(
v_{i}^{n,j}\right) ^{2}h\right) ^{1/2}$

The scheme converges when the norm of that error $\left\|
V^{j}\right\| \rightarrow 0$ as $\left( \tau ,\text{ }h\text{
}\rightarrow 0\right) $\\
We found that \ $\left\| V^{j+1}\right\| \leq \,P(M)\,\ O\left(
\tau +h^{2}\right) $, where $\,P(M)\,\ $is a polynomial in the
bounded constant $M=\tau\frac{e^{a \tau j}-1}{e^{a\tau}-1}$ and
$a$ as in (\ref{a}). Hence the convergence proved.

\subsection{Numerical calculations and test}

The coupled KdV-MKdV system (\ref{gen-sys2}) is solved
numerically using scheme (\ref{schem}) with initial condition
from (\ref{sol-gen-sys2}) at $t=0
$ and the results are compared with the explicit formulas (\ref{sol-gen-sys2}%
). The percentage errors are shown in the following plots.

\begin{center}
\epsfig{file=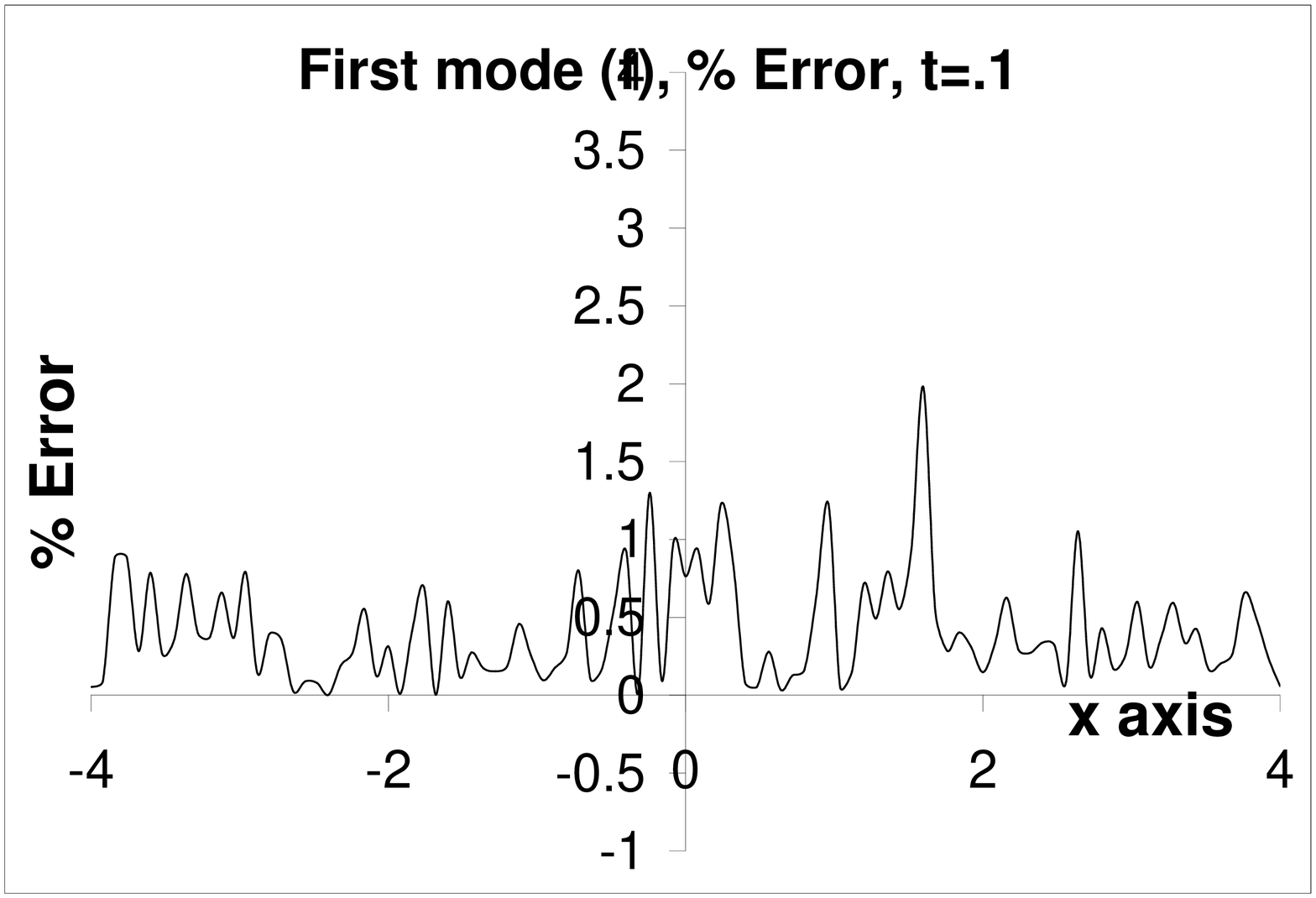, height=3.5  cm, width=4.5 cm ,clip= ,
angle=0} \epsfig{file=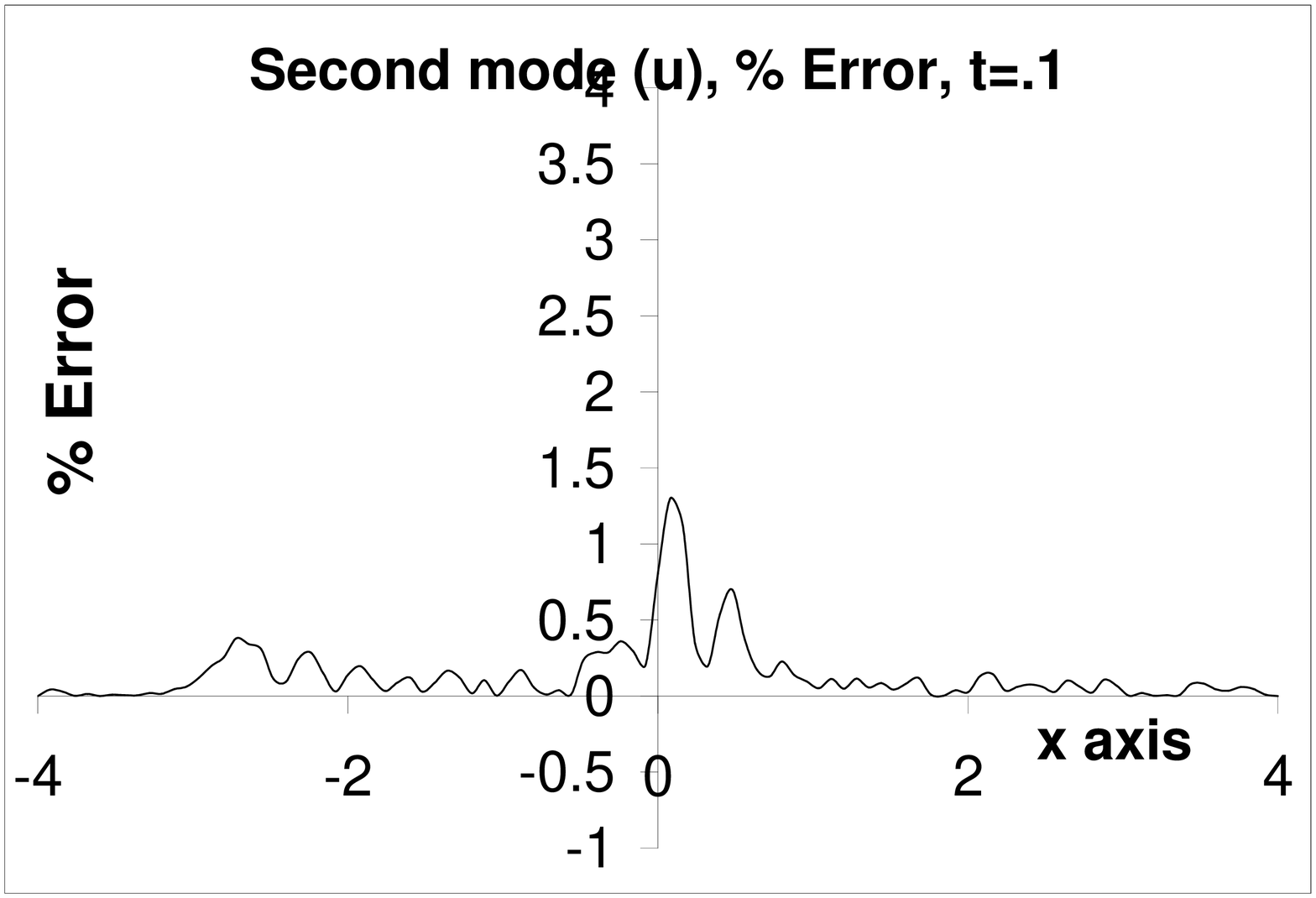,height=3.5 cm, width=4.5cm
,clip=,angle=0} \epsfig{file=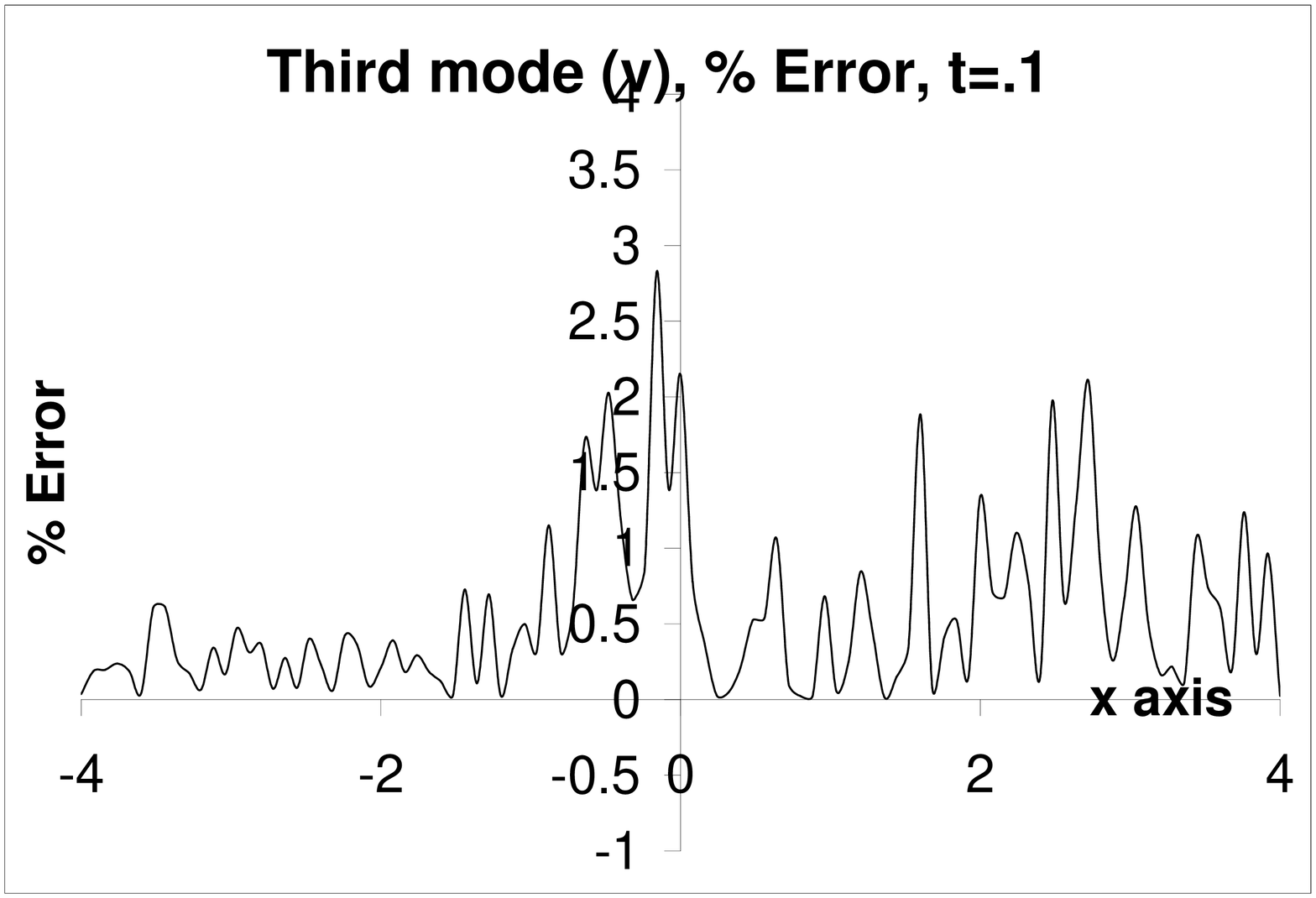,height=3.5 cm,
width=4.5cm
,clip=,angle=0}\\[0pt]
\ Fig.3 percentage errors of the numerical solutions relative  to
the explicit solutions.
\end{center}
The results of the test confirms the validity of the numerical
scheme we propose. It also illustrates the errors of evaluation
that could be estimated by the resulting inequalities of the
scheme convergence proof.

\section{Conclusion}
Darboux transformations covariance of Lax representation of
nonlinear equations is a powerful tool for explicit solutions
production. Here we investigate applications of special kind of
such discrete symmetry - to be called elementary ones. We use
these elementary DT to produce explicit solutions for coupled
KdV-MKdV system. The iteration of DT can be formulated in form of
determinants representations \cite{Leb-Ust,Leb-Ust2}. A numerical
method for general coupled KdV-MKdV system is introduced. It is a
difference scheme for Cauchy problems for arbitrary number of
equations with constants coefficients. We analyze stability and
prove the convergence of the scheme. The scheme keeps two
conservation laws chosen in analogy with KdV type equations.
Analyzing stability and proving the convergence beside comparing
the numerical results with explicit formulas allow us to use the
numerical scheme to systems with arbitrary coefficients that is
presumably non-integrable. Obviously the coupled KdV systems are
successfully treated by our scheme \cite{Halim}. \\

\begin{Large}{Acknowledgment}
\end{Large} We thank S. B. Kshevetskii for useful discussions about numerical
scheme for the problem under consideration.

\section{Appendix A\newline
Stability analysis of the scheme}

We prove stability with respect to small perturbations (because we
consider nonlinear equations) of initial conditions. Strictly
speaking it is the boundness of the discrete solution in terms of
small perturbation of the initial data. Consider the differential

$T_{i,r}^{n,j+1}=\left\{ \partial \theta _{i}^{n,j+1}/\partial
\theta _{r}^{n,j}\right\} \,,\,d\theta _{r}^{n,j}=\left\{ \theta
_{i-2}^{n,j}\,\ \theta _{i-1}^{n,j}\,\ \theta _{i}^{n,j}\,\
\theta _{i+1}^{n,j}\,\ \theta _{i+2}^{n,j}\right\} ^{t}$ and
define the norm $\left\| d\theta ^{j}\right\| =\left(
\underset{r}{\sum }\underset{n}{\sum }\left( d\theta
_{r}^{n,j}\right) ^{2}\,\ h\right) ^{1/2}$

We can write \ $d\theta _{i}^{n,j+1}=T_{i,r}^{n,j+1}\,d\theta
_{r}^{n,j}=T_{i,r}^{n,j+1}\,T_{i,r}^{n,j}\,d\theta _{r}^{n,j-1}=\underset{r}{%
\Pi }\,\left( T_{i,r}^{n}\right) ^{r}\,d\theta _{r}^{n,o}$

where $d\theta _{i}^{n,j+1}$ is perturbations of the discrete solution, $%
d\theta _{r}^{n,o}$ small perturbation of the initial data.
Stability required the boundedness of \thinspace
$\underset{r}{\Pi }\,\left( T_{i,r}^{n}\right)
^{r}\,\,$i.e\thinspace\ $\left\| T^{r}\right\| \,$is bounded. We
calculate $T\,\,\ $from (\ref{schem}) as follow
\begin{gather}\label{T}
T_{i,r}^{n,j+1}=\delta _{i,r}- \tau \underset{m,k}{\sum }( \frac{g_{m,k}^{n,1}%
}{2h}\left[ \theta _{i}^{m,j}\left( \delta _{i+1,r}-\delta
_{i-1,r}\right)
+\delta _{i,r}\left( \theta _{i+1}^{k,j}-\theta _{i-1}^{k,j}\right) \right] %
\notag   \\
+\frac{g_{m,k}^{n,2}}{2h}\left[ \left( \theta _{i}^{m,j}\right)
^{2}\left( \delta _{i+1,r}-\delta _{i-1,r}\right) +2\theta
_{i}^{m,j}\delta
_{i,r}\left( \theta _{i+1}^{k,j}-\theta _{i-1}^{k,j}\right) \right] \notag  \\
+\frac{g_{m,k}^{n,3}}{2h}\left[ \left( \theta _{i+1}^{m,j}-\theta
_{i-1}^{m,j}\right) \left( \delta _{i+1,r}-\delta _{i-1,r}\right)
+\left( \delta _{i+1,r}-\delta _{i-1,r}\right) \left( \theta
_{i+1}^{k,j}-\theta
_{i-1}^{k,j}\right) \right] \notag  \\
+\frac{g_{m,k}^{n,4}}{2h}\left[ \theta _{i}^{m,j}\left( \delta
_{i+1,r}-2\delta _{i,r}+\delta _{i-1,r}\right) +\delta
_{i,r}\left( \theta
_{i+1}^{k,j}-2\theta _{i}^{k,j}+\theta _{i-1}^{k,j}\right) \right] \notag  \\
+\frac{g_{m,k}^{n,5}}{2h}\left[ \theta _{i}^{m,j}\theta
_{i}^{k,j}\left( \delta _{i+1,r}-\delta _{i-1,r}\right) +\theta
_{i}^{m,j}\left( \theta _{i+1}^{k,j}-\theta _{i-1}^{k,j}\right)
\delta _{i,r}+\theta
_{i}^{k,j}\left( \theta _{i+1}^{k,j}-\theta _{i-1}^{k,j}\right) \delta _{i,r}%
\right] \notag ) \\
-\frac{\tau d_{n}}{2h^{3}}\left[ \delta _{i+2,r}-2\delta
_{i+1,r}+2\delta _{i-1,r}-2\delta _{i-2,r}\right] \label{T}
\end{gather}
Rewriting (\ref{T}) in terms of  identity (E), symmetric (S) and
anti-symmetric (A) matrices\newline

$\left\| S^{j+1}\right\| \leq \tau \,\underset{l,n,m,k}{\max
}\left| g_{m,k}^{n,l}\right| \,\ \underset{i,m,k}{\max }\left(
\left| \theta _{x,i}^{m,j}\right| \,\left| \theta
_{x,i}^{k,j}\right| \right) ,$

$\left\| A^{j+1}\right\| \leq \frac{\tau }{h}\,\underset{l,n,m,k}{\max }%
\left| g_{m,k}^{n,l}\right| \,\ \underset{i,m,k}{\max }\left(
\left| \theta
_{i}^{m,j}\right| \,\left| \theta _{i}^{k,j}\right| \right) +\frac{3\,}{h^{3}%
}\underset{n}{\max }\left| d_{n}\right| ,$

where $\theta _{x,i}^{j}=\frac{\theta _{i+1}^{j}-\theta
_{i-1}^{j}}{2h}\,\ ,\,\ n,m,k=1,2,..N,\,\ l=1,2,...5.$\newline

$\left\| T^{j+1}\right\| ^{2}=\left\| \left( T^{j+1}\right)
^{\ast }\,\ T^{j+1}\right\| =\left\| \left(
E-A^{j+1}+S^{j+1}\right) \left( E+A^{j+1}+S^{j+1}\right) \right\|
$

\ \ \ \ \ \ \ \ \ \ \ \ \ \ \ $\leq 1+2\,\left\| S^{j+1}\right\|
+\left( \left\| A^{j+1}\right\| +\left\| S^{j+1}\right\| \right)
^{2}$

\thinspace\ \ \ \ \ \ \ \ \ \ \ \ \ \ \ $\leq e^{a(\tau, h)\tau
},$

\begin{gather*}
a(\tau, h) = 2 \underset{l,n,m,k}{\max }\left|
g_{m,k}^{n,l}\right| \,\ \underset{ i,m,k}{\max }\left( \left|
\theta _{x,i}^{m,j}\right| \,\left| \theta _{x,i}^{k,j}\right|
\right) +\tau [ \underset{l,n,m,k}{\max }\left|
g_{m,k}^{n,l}\right| \,\ \underset{i,m,k}{\max }\left( \left|
\theta
_{x,i}^{m,j}\right| \,\left| \theta _{x,i}^{k,j}\right| \right)    \\
\hspace*{10 mm}+\frac{1}{h} \,\underset{l,n,m,k}{\max }\left|
g_{m,k}^{n,l}\right| \,\ \underset{i,m,k}{ \max }\left( \left|
\theta _{i}^{m,j}\right| \,\left| \theta _{i}^{k,j}\right| \right)
+\frac{3\,}{h^{3}}\underset{n}{\max }\left| d_{n}\right| ]^{2}
\end{gather*}
We have here a conditional stability. That is we require that
$\tau \rightarrow 0\,\ $ more faster than $h\rightarrow
0.\,$Namely we need
\begin{gather*}
  \tau \leq \,\left( \text{cons}\tan
\text{t}\right) \,.\,h^{6}
\end{gather*}
\section{Appendix B\newline
The scheme convergence}

We prove the convergence by proving that the norm of the error
(between the difference solution and the exact solution) vanishes
as the mesh is refined. Let $\theta _{i}^{j}\,$the difference
solution of (\ref{schem}), u$_{i}^{j}\, $the exact solution. The\
error v$_{i}^{j}$ is given by $v_{i}^{j}=\theta
_{i}^{j}-u_{i}^{j}\,.\,$ \newline
The scheme converges when the norm $\left\| V^{j}\right\| \rightarrow 0$ as $%
\left( \tau ,\text{ }h\text{ }\rightarrow 0\right) $ where the
norm is defined as $\left\| V^{j}\right\| =\left(
\underset{i}{\sum }\underset{n}{\sum }\left( v_{i}^{n,j}\right)
^{2}h\right) ^{1/2}$

substitute in (\ref{schem}) by \ $\theta _{i}^{j}=v_{i}^{j}+u_{i}^{j}\,\ $%
keeping in mind that for $\theta _{i}^{j}\,$equation (\ref{schem}) is $%
O\,\left( \tau +h^{2}\right) \,$and using the operator T defined
by\newline
\begin{gather*}
v_{i}^{n,j}-\tau \lbrack \underset{m,k}{\sum }(g_{m,k}^{n,1}(
u_{i}^{m,j}\frac{v_{i+1}^{k,j}-v_{i-1}^{k,j}}{2h}+v_{i}^{m,j}\frac{
u_{i+1}^{k,j}-u_{i-1}^{k,j}}{2h}) +g_{m,k}^{n,2}(
(u_{i}^{m,j})^{2}\frac{v_{i+1}^{k,j}-v_{i-1}^{k,j}}{2h} \\
+(v_{i}^{m,j})^{2}\frac{u_{i+1}^{k,j}-u_{i-1}^{k,j}}{2h})
+g_{m,k}^{n,3}( \frac{u_{i+1}^{m,j}-u_{i-1}^{m,j}}{2h}\frac{
v_{i+1}^{k,j}-v_{i-1}^{k,j}}{2h}+\frac{v_{i+1}^{m,j}-v_{i-1}^{m,j}}{2h}\frac{
u_{i+1}^{k,j}-u_{i-1}^{k,j}}{2h}) \\
+g_{m,k}^{n,4}( u_{i}^{m,j}
\frac{v_{i+1}^{k,j}-2v_{i}^{k,j}+v_{i-1}^{k,j}}{2h}+v_{i}^{m,j}\frac{
u_{i+1}^{k,j}-2u_{i}^{k,j}+u_{i-1}^{k,j}}{2h})
+2g_{m,k}^{n,5}( u_{i}^{m,j}v_{i}^{k,j}\\
\frac{ v_{i+1}^{k,j}-v_{i-1}^{k,j}}{2h}
+v_{i}^{m,j}u_{i}^{k,j}\frac{ u_{i+1}^{k,j}-u_{i-1}^{k,j}}{2h}) )
+d_{n}\frac{
v_{i+2}^{n,j}-2v_{i+1}^{n,j}+2v_{i-1}^{n,j}-v_{i-2}^{n,j}}{2h^{3}}]\\
\hspace{5 mm}= \underset {r}{\sum }T_{ir}^{j+1}v_{r}^{n,j}
\end{gather*}
\newline So we obtain
\begin{equation}  \label{error}
v_{i}^{n,j+1}=\underset{r}{\sum }T_{ir}^{j+1}v_{r}^{n,j}+\tau
\,f_{m,k,i}^{n,j}  
\end{equation}
where $f_{m,k,i}^{n,j}=\underset{m,k}{\sum }g_{m,k}^{n,1}v_{i}^{m,j}\frac{%
v_{i+1}^{k,j}-v_{i-1}^{k,j}}{2h}+g_{m,k}^{n,2}(v_{i}^{m,j})^{2}\frac{%
v_{i+1}^{k,j}-v_{i-1}^{k,j}}{2h}+g_{m,k}^{n,3}\frac{%
v_{i+1}^{m,j}-v_{i-1}^{m,j}}{2h}\frac{v_{i+1}^{k,j}-v_{i-1}^{k,j}}{2h}  \\
\hspace*{15 mm}+g_{m,k}^{n,4}v_{i}^{m,j}\frac{v_{i+1}^{k,j}-2v_{i}^{k,j}+v_{i-1}^{k,j}}{2h}%
 +2g_{m,k}^{n,5}v_{i}^{m,j}v_{i}^{k,j}\frac{%
v_{i+1}^{k,j}-v_{i-1}^{k,j}}{2h}+O\left( \tau +h^{2}\right)
$\newline

$\left\| f^{j}\right\| =\left( \underset{i}{\sum }\left(
f_{m,k,i}^{n,j}\right) ^{2}h\right) ^{1/2}$\newline
\hspace*{11 mm} \ $\leq \frac{\left| g_{m,k}^{n,1}\right| _{\max }}{h^{3/2}}%
\left\| V^{j}\right\| ^{2}+\frac{\left| g_{m,k}^{n,2}\right| _{\max }}{h^{2}}%
\left\| V^{j}\right\| ^{3}+\frac{\left| g_{m,k}^{n,3}\right| _{\max }}{%
h^{5/2}}\left\| V^{j}\right\| ^{2}+\frac{\left| g_{m,k}^{n,4}\right| _{\max }%
}{h^{5/2}}\left\| V^{j}\right\| ^{2}$\newline
$\hspace*{14 mm}+\frac{\left|g_{m,k}^{n,5}\right| _{\max }} {h^{2}}%
 \left\|V^{j}\right\| ^{3}+O\left( \tau +h^{2}\right) $

\hspace*{8 mm} $\leq \frac{\left| g_{m,k}^{n,l}\right| _{\max }}{h^{2}}%
\left\| V^{j}\right\| ^{3}+O\left( \tau +h^{2}\right),
\hspace*{10 mm}
\left| g_{m,k}^{n,l}\right| _{\max }=\underset{n,m,k,1}{\max }%
g_{m,k}^{n,l}$\newline

Using Schwartz inequality so (\ref{error}) becomes

$\left\| V^{\ j+1}\right\| \leq \left\| T^{j+1}\right\| \left\|
V^{j}\right\| +\tau \left\| f^{j}\right\| $

\hspace*{13 mm} $\leq \left\| T^{j+1}\right\| \left\|
T^{j}\right\| \left\| V^{j-1}\right\| +\tau \left( T^{j+1}\left\|
f^{j-1}\right\| +\left\| f^{j}\right\| \right) $

\hspace*{13 mm} $\leq e^{a\tau j}\left\| V^{o}\right\| +\tau
\left( e^{a\tau \left( j-1\right) }\left\| f^{o}\right\| +e^{a\tau
\left( j-2\right) }\left\| f^{1}\right\| +...+\left\|
f^{j}\right\| \right) $

\hspace*{13 mm} $\leq e^{a\tau j}\left\| V^{o}\right\| +M\,\ \
\left| g_{m,k}^{n,l}\right| _{\max }\left\| V^{j+1}\right\|
^{3}+M\,\ O\left( \tau +h^{2}\right), M=\tau\frac{e^{a \tau
j}-1}{e^{a\tau}-1} $\newline

Using $\left\| V^{o}\right\| =0,$ the above inequality has the
solution

$\left\| V^{j+1}\right\| \leq P\left( M\right) \,\ \ O\,\left(
\tau +h^{2}\right) ,\,\ P(M)$ is a polynomial in $M$.\\
 Since $M$
is bounded then $\left\| V^{j+1}\right\| \rightarrow 0\,\ as\,\
\tau ,\,h\,\rightarrow 0\,\ $and the convergence proved.

\end{document}